\documentclass[format=sigconf,anonymous=false,review=false]{acmart}

\renewcommand\footnotetextcopyrightpermission[1]{} 

\AtBeginDocument{%
  \providecommand\BibTeX{{%
    \normalfont B\kern-0.5em{\scshape i\kern-0.25em b}\kern-0.8em\TeX}}}

\setcopyright{acmcopyright}
\copyrightyear{2022}
\acmYear{2022}
\acmDOI{nnnnn}

\acmConference[Under review]{ }{2022}{Preprint}
\acmBooktitle{ Under review}
\acmPrice{15.00}
\acmISBN{nnn}

\acmSubmissionID{3098}

\usepackage{booktabs} 
\usepackage{diagbox} 
\usepackage{graphicx}
\usepackage[caption=false]{subfig}
\usepackage{color}
\usepackage{microtype}
\usepackage{amsmath,amsfonts}
\usepackage{mathtools}  
\usepackage{mathrsfs} 
\usepackage{bm}
\usepackage{booktabs, multirow}  
\usepackage{makecell}
\usepackage{stfloats}

\begin{document}

\title{AdaSparse: Learning Adaptively Sparse Structures for Multi-Domain Click-Through Rate Prediction}

\author{Xuanhua Yang, Xiaoyu Peng, Penghui Wei$^{\scriptscriptstyle *}$, Shaoguo Liu$^{\scriptscriptstyle *}$, Liang Wang and Bo Zheng}
\thanks{${\scriptscriptstyle *}$ Correspondence to: S. Liu and P. Wei.}
\affiliation{
  \institution{Alibaba Group} 
  \country{Beijing, China}
}
\email{{xuanhua.yxh,pengxiaoyu.pxy,wph242967,shaoguo.lsg,	liangbo.wl,bozheng}@alibaba-inc.com}

\begin{abstract}
Click-through rate (CTR) prediction is a fundamental technique in recommendation and advertising systems. Recent studies have proved that learning a unified model to serve multiple domains is effective to improve the overall performance. However, it is still challenging to improve generalization across domains under limited training data, and hard to deploy current solutions due to computational complexity. In this paper, we propose \textsf{AdaSparse} for multi-domain CTR prediction, which learns adaptively sparse structure for each domain, achieving better generalization across domains with lower computational cost. We introduce domain-aware neuron-level weighting factors to measure the importance of neurons, with that for each domain our model can prune redundant neurons to improve generalization. We further add flexible sparsity regularizations to control the sparsity ratio of learned structures. Offline and online experiments show that \textsf{AdaSparse} outperforms previous multi-domain CTR models significantly.
\end{abstract}

\maketitle
\fancyhead{}

\section{Introduction}\label{intro}
\begin{figure}[t]
\centering
\centerline{\includegraphics[width=1.0\columnwidth]{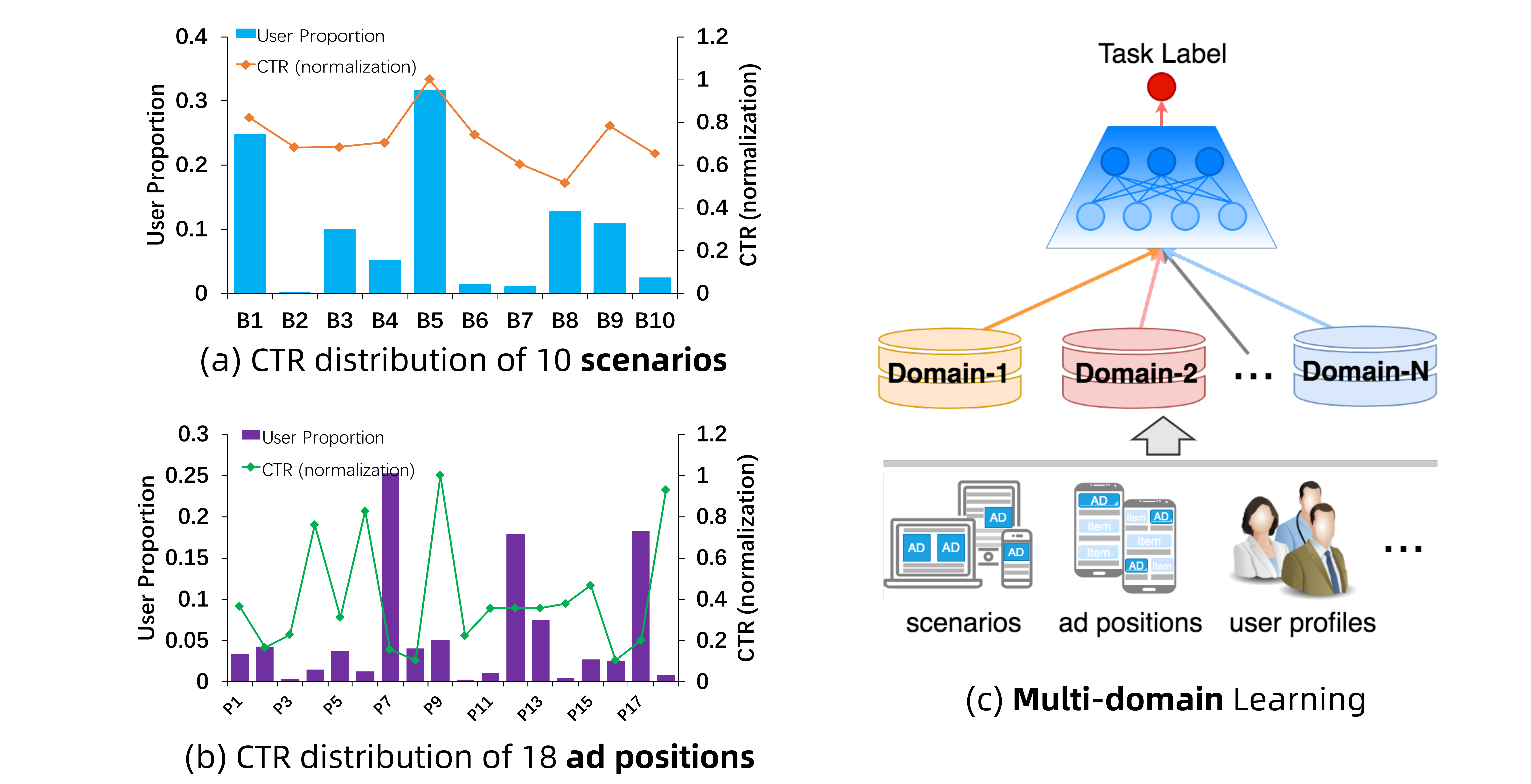}}
\vspace{-1em}
\caption{Two cases of domain differences are shown in (a) and (b). Multi-domain learning paradigm is shown in (c).}
\label{fig:multi-domain-pic}
\vspace{-1em}
\end{figure}

In online advertising and recommendation systems, click-through rate (CTR) prediction is a fundamental technique. There is a great need to capture differentiated user preferences for \textit{multiple domains}. 
On one hand, there are always more than one \textit{scenario} (e.g., search, feeds and others) that require the ability of prediction models, however their data are usually not homologous. Fig.~\ref{fig:multi-domain-pic} (a) shows the CTR distribution of 10 business scenarios from our advertising platform, and we can see that the averaged CTR from large/small scenarios (e.g., B5/B2) are quite different. 
On the other hand, if we partition the data with \textit{representative features} such as the position of displayed ads, from Fig.~\ref{fig:multi-domain-pic} (b)  we can also observe that the subsets with different feature values have different natures. 

As in Fig.~\ref{fig:multi-domain-pic} (c), recent studies pay more attention to multi-domain learning paradigm in CTR prediction, aiming to improve the overall performance on all domains. 
There are two crucial issues: 
(1) \textbf{Generalization} across domains. To achieve better effectiveness, the model need to possess the ability of capturing the user interest specific to each domain. Besides, advertisers often change delivery strategies (such as delivery scenarios, targeted audiences and ad positions), leading to emerging domains having insufficient training data. 
(2) \textbf{Computational cost}. Considering the online deployment procedure, the model should be parameter-efficient, contributing to memory space and response time in online service. 

Recent studies in multi-domain CTR prediction can be divided into two categories. 
(1) Models with \textbf{individual parameters}~\cite{yu2020personalized,sheng2021one}, in which each domain has its separate learnable parameters.~\citet{yu2020personalized} regard each domain as a meta-task and learn task-specific parameters via meta-learning.~\citet{sheng2021one} decompose all parameters to shared part and domain-specific part, where each domain's parameters are trained using the data from this domain only. They are usually inefficient on computation and memory, because the complexity grows much faster with the increase of domains. Besides, the generalization for domains having limited training data is not ideal. 
(2) Models with \textbf{generated parameters}~\cite{yan2022apg}. An auxiliary network is used to dynamically generate domain-specific parameters, taking domain-aware features as input. Due to the large amount of parameters to be generated in current solutions, it makes the models hard to converge and does harm to generalization. These models are also faced with computational cost issue. 

We propose a simple yet effective framework \textsf{AdaSparse}, which learns adaptively sparse structure for each domain to take both the generalization across domains and computational efficiency into account. 
We introduce domain-aware neuron-level weighting factors to measure the importance of neurons w.r.t. different domains. With that our model can prune redundant neurons to improve generalization, especially for domains with limited training data. We perform neuron-level rather than connection-level pruning to guarantee lower computation complexity, because the number of neurons is far less than connections. 
Our contributions are:
\begin{itemize}
    \item To our knowledge, this is the first work that learns sparse structures to take both generalization and computational cost into account for multi-domain CTR prediction.  
    \item We propose \textsf{AdaSparse} to adaptively prune redundant neurons w.r.t. different domains via learned neuron-level weighting factors to improve generalization, which also guarantees lower computation complexity. 
    \item Both offline and online experiments verify that \textsf{AdaSparse} significantly outperforms previous models. We show that the learned sparse structures capture domain distinction.
\end{itemize}

\begin{figure}[t]
\centering
\centerline{\includegraphics[width=0.9\columnwidth]{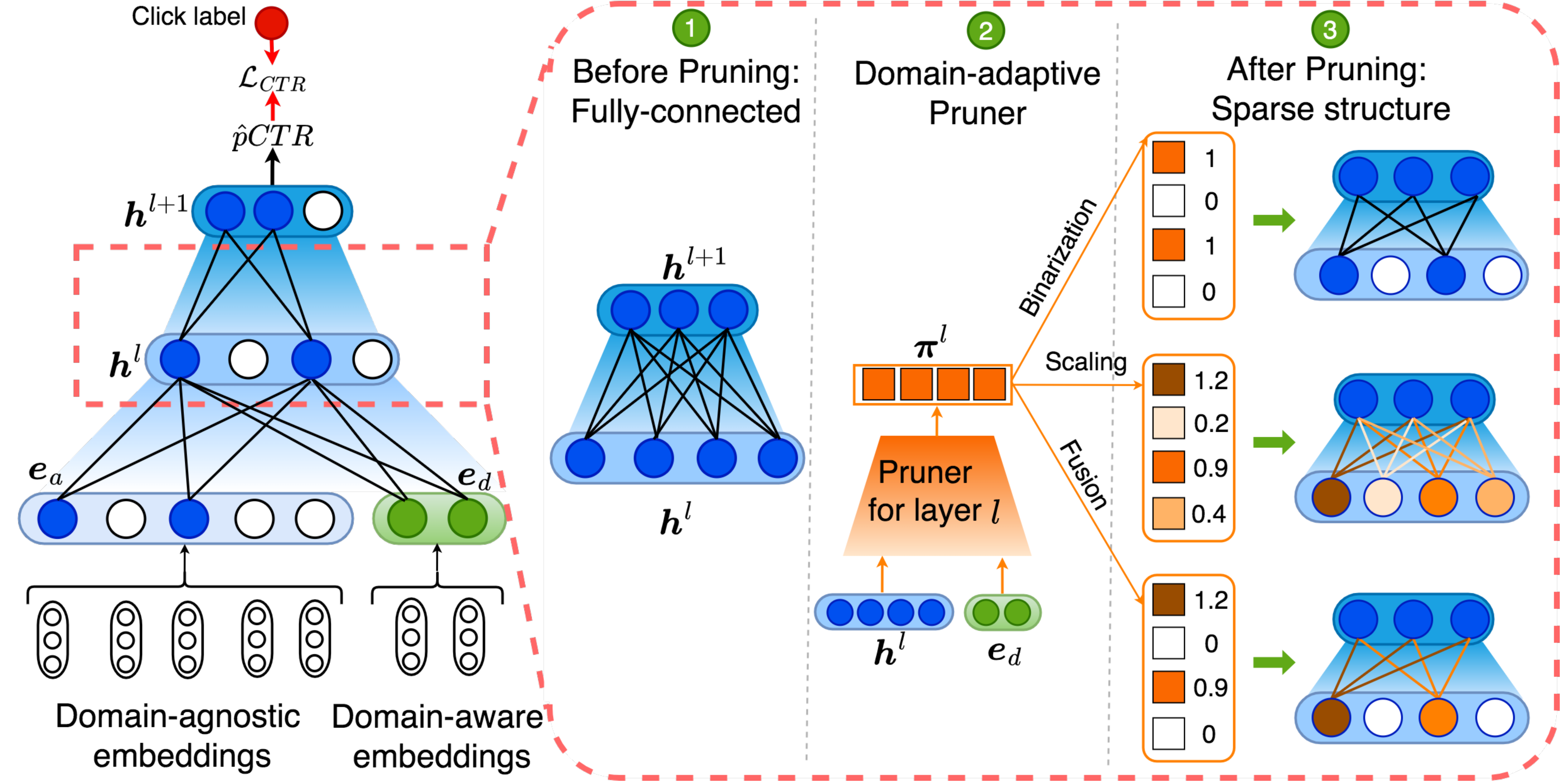}}
\vspace{-1em}
\caption{\textsf{AdaSparse} for multi-domain CTR prediction.}
\label{fig:overview}
\vspace{-1em}
\end{figure}

\section{Problem Formulation}
Considering a training dataset $\mathcal{D} = \{(x_j,y_j)\}^{|\mathcal D|}_{j=1}$, where $x_j$ and $y_j$ represent the feature set and binary click label of the $j_{th}$ sample. 

We partition the dataset $\mathcal{D}$ to multiple domain-specific subsets $\mathcal D^d$ (that is, $\mathcal{D}=\bigcup_{d} \mathcal{D}^d$), where domain $d$'s subset $\mathcal{D}^d = \left\{\left(x^d_i,x^{a}_i,y_i\right)\right\}^{|\mathcal D^d|}_{i=1}$ is obtained based on \textit{domain-aware feature set}: here, the whole feature set $x_i$ is divided by domain-aware feature set $x^d_i$ and domain-agnostic feature set $x^{a}_i$. Take Fig.~\ref{fig:multi-domain-pic} (b) as an example, the domain-aware feature set $x^d_i$ contains one feature {\small \texttt{ad\_position}}, resulting in 18 domains in $\mathcal{D}$ and the data sizes of these domains are usually quite different. 
We can also assign more features to construct the domain-aware feature set, such as {\small$\{\texttt{scenario}, \texttt{user\_profile}, \texttt{ad\_position}\}$}, and this may result in thousands of domains. 
The goal of multi-domain CTR prediction is to learn a  model $\hat{p}CTR=f(x)$ that performs well on all domains.

\section{Proposed Approach}
We propose \textsf{AdaSparse} that learns adaptively sparse structure for each domain to achieve better generalization. Figure~\ref{fig:overview} givens an overview. The core is a domain-aware pruner that produces neuron-level weighting factors to prune redundant neurons.

Without loss of generality, we use an $L$-layered fully-connected neural network~\cite{covington2016deep} as the backbone model architecture $f(\cdot)$ to introduce our \textsf{AdaSparse}. Our model operates on neuron-level, thus it can be easily extended to other common architectures in CTR prediction, such as deep\&cross network~\cite{wang2021dcn}. 

After transforming all features into embeddings, we concatenate domain-aware / -agnostic feature embeddings as $\bm e_d$ / $\bm e_a$. The model input is the concatenation of $[\bm e_d,\bm e_a]$. 
Let $\textbf{W}^l$ denote learnable matrix of the $l_{th}$ layer. That is, $\bm{h}^{l+1}=\tanh\left(\textbf{W}^l\bm{h}^l\right)$, where $\bm{h}^{l}$ is the input (neuron) of the $l_{th}$ layer and we omit the bias for simplification. The model is trained through cross-entropy criterion $\mathcal{L}_{CTR}(y,\hat{p}CTR)$ over all domains' samples:
\begin{equation}\label{eq:optimization0}
\min_{\textbf{W}} \frac{1}{|\mathcal D|}\sum_{d}\sum_{i=1}^{|\mathcal D^d|}\mathcal{L}_{CTR}\left(y_i,f\left({x}^d_i,x^{a}_i\right);\{\textbf{W}^l\}_{1 \leq l \leq L }\right)\,.
\end{equation}

For given domain $d$, \textsf{AdaSparse} employs the domain-aware pruner to adaptively remove each layer's neurons. 
Specifically, consider the $l_{th}$ layer's matrix $\textbf{W}^l\in\mathbb R^{N_l\times N_{l+1}}$ that connects from $N_l$ neurons $\bm{h}^{l}$ to next layer's $N_{l+1}$ neurons $\bm{h}^{l+1}$, the domain-aware pruner produces a weighting factor vector $\bm{\pi}^l(d)\in\mathbb R^{N_l}$ to prune the $\bm{h}^{l}$. It operates for each layer and finally we obtain a sparse structure.

\subsection{Domain-adaptive Pruner}\label{method:Adapter}
\subsubsection{\textbf{Lightweight Pruner}}\label{method:domain-specific-factor} 

As in the right part of Fig.~\ref{fig:overview}, the domain-aware pruner is a lightweight network that takes $\bm{h}^{l}$ and domain-aware features $\bm e_d$ as input. 

For each layer $l$, it outputs the neuron-level weighting factors $\bm{\pi}^l(d)$, which is sparse and used to prune the neurons by
\begin{equation}
    \bm h^l(d) \coloneqq \bm h^l \odot \bm{\pi}^l(d)
\end{equation}
where $\odot$ is element-wise multiplication. We use sigmoid $\sigma(.)$ as activation function of the pruner, and employ a soft-threshold operator $S_{\epsilon}(.)$ to enforce that the outputs are less than a predefined value $\epsilon$. Formally, the computation for the $l_{th}$ layer of pruner is:
\begin{equation}\label{eq:adapter}
\bm{\pi^l}(d)=S_{\epsilon}\left(\sigma\left(\textbf{W}^l_{p} \cdot [\bm{e}_d;\bm{h}^l]\right)\right),
\end{equation}
where $\textbf{W}^l_{p}$ denotes the learnable parameters of the $l_{th}$ layer.

\subsubsection{\textbf{Formulations of Weighting Factors}}\label{method:three-types-factor} 
We introduce three formulations of weighting factors $\bm{\pi}^l$, and detail the subtle difference about their sigmoid function and soft-threshold operator, inspired by model compression techniques~\cite{liu2017learning,zhu2017prune,luo2020autopruner}.

1. \textbf{Binarization} method. It requires that all elements of $\bm{\pi}^l$ should be 0-1 binary value. It is challenging to achieve this goal automatically if we only use original sigmoid function $v_{out}=\sigma(v_{in})$, because the element $v_{out}$ can only be approximate 0 or 1 value when $|v_{in}|$ becomes magnitude. 
Or if we use an improper threshold (e.g. $0.5$) to truncate the output value $v_{out}$, we may prune the neurons by accident when the model is still under-fitting during early training. 

We incorporate a parameter $\alpha$ into the sigmoid function:
\begin{equation}
v_{out}=\sigma(\alpha \cdot v_{in})    
\end{equation}
and gradually increase $\alpha$ during training. When $\alpha$ is large enough, $v_{out}$ will converge to 0-1 value. 
Here, we formulate the soft-threshold operator $S_{\epsilon}(.)$ as: 
\begin{equation}
    S_{\epsilon}(v_{out})=sign(|v_{out}|-\epsilon)
\end{equation}
where $\epsilon > 0$ is a tiny threshold.\footnote{$sign(\cdot)$ is short for signum function, $sign(x)=\left\{
\begin{array}{cl}
1 &  x > 0 \\
0 &  x \leq 0 \\
\end{array} \right.$} 
In binarization method, the prune operation $\bm h^l(d) \coloneqq \bm h^l \odot \bm{\pi}^l(d)$ can be implemented using sparse operators to accelerate the computation.\footnote{Such as \texttt{tf.sparse.sparse\_dense\_matmul} in TensorFlow.}

2. \textbf{Scaling} method. We regard binarization method as a hard weighting, and here Scaling method is a soft one: 
\begin{equation}
v_{out}=\beta\cdot\sigma( v_{in}),\quad \beta \geq 1
\end{equation}
which produces factor values ranging from 0 to $\beta$. It is expected that the more important neurons, the larger factor values they will have. We design the soft-threshold operator of {Scaling} method as
\begin{equation}
    S_{\epsilon}(v_{out}) = v_{out} \cdot sign(|v_{out}|)\,.
\end{equation}

3. \textbf{Fusion} method. In Scaling method we give less important neurons very small factor values. To prune the redundant information as much as possible, we combine {Binarization} and {Scaling} methods by formulating:
\begin{equation}
\begin{aligned}
v_{out}&=\beta\cdot\sigma(\alpha\cdot v_{in}),\quad \beta \geq 1\\
S_{\epsilon}(v_{out}) &= v_{out} \cdot sign(|v_{out}|-\epsilon),\quad \epsilon >0\,.
\end{aligned}
\end{equation}

\subsubsection{\textbf{Discussion on Memory and Computational Costs}}
As we mentioned in §~\ref{intro}, there are two types of approaches for multi-domain CTR prediction, based on individual parameters and generated parameters respectively. \textsf{AdaSparse} belongs to the latter. 

\textsf{AdaSparse} only incorporates an additional network as pruner, thus it is parameter-efficient and the memory cost is much smaller than individual parameters. In terms of computational cost, we compare it with individual-based \textsf{STAR}~\cite{sheng2021one} and generated-based \textsf{APG}~\cite{yan2022apg}. Let $D$ denote the number of domains, and take the computation for $l_{th}$ layer for comparison. The complexities of \textsf{STAR}, \textsf{APG} and \textsf{AdaSparse} are $\mathcal{O}(D\cdot N_l \cdot N_{l+1})$, $\mathcal{O}(D\cdot ( N_l K+ N_{l+1} K))$ and $\mathcal{O}(D\cdot N_l)$ respectively, verifying that our approach is time-efficient. 

\subsection{Sparsity Controlled Regularization}\label{method:Regularization}

In Binarization method, a core problem is that we need to control the weighting factors of redundant neurons gradually converge to zero during training. 
We propose a flexible sparsity regularization to control learning of factors $\bm{\pi}^l(d)$, and it also meets the various requirements of sparsity ratio for learned structures.

In {Binarization} method, $\bm{\pi}^l$ is a 0-1 binary vector, and an usual sparsity regularization way is adding $\ell_1$-norm term $||\bm{\pi}^l||_1$. However, we empirically find that it is hard to achieve our expected sparsity ratio without explicit controlling. 

Considering that $\frac{{||\bm{\pi}^l||_1}}{N_l}$ represents the percentage of 1 code in $\bm{\pi}^l$, we denote the sparsity ratio as $r^l=1-\frac{{||\bm{\pi}^l||_1}}{N_l}$. 
We then predefine \textit{sparsity ratio boundary} $[r_{min},r_{max}]$ as our expected goal, and explicit control the actual sparsity ratio during training by adding an sparsity regularization loss term: 
\begin{equation}\label{eq:sparsity_reg}
R_s\left(\{\bm{\pi}^l\}_{1 \leq l \leq L }\right)=\frac{1}{L}\sum_{l=1}^{L}\lambda^{l}||r^l-r||_2, \quad r=(r_{min} + r_{max}) / 2,
\end{equation}
where $\lambda^l$ is an dynamic balancing weight based on the current actual sparsity rate $r^l$ in training procedure:
\begin{equation}\label{eq:sparsity_reg}
\lambda^{l}=\left\{
\begin{array}{cl}
0 &  r^l \in \left[r_{min},r_{max}\right]  \\
\hat{\lambda} \cdot |r^l-r| &  r^l\notin \left[r_{min},r_{max}\right] \\
\end{array} \right.,
\end{equation}
where $\hat{\lambda}$ is initialized by 0.01 and gradually increase over training step, thus the model pays less attention on sparsity loss term during early training and focuses more on adjusting $\bm{\pi}^l$.
We can see that when the current sparsity ratio $r^l$ falls into our expected boundary $\left[r_{min},r_{max}\right]$, $\lambda ^l=0$ and this loss term will not affect the learning.

\section{Experiments}

\subsection{Experimental Setup}
\subsubsection{\textbf{Datasets}} We conduct experiments on two datasets. One is a million-scale public dataset called \textbf{IAAC}~\cite{yan2022apg} that has 10.9 million impressions and more than \textit{300 domains}. Another is a billion-scale production dataset called \textbf{Production} that was collected from our advertising system and has 2.2 billion impressions and about \textit{5,000 domains}.\footnote{Selected domain-aware feature set: \{\texttt{user\_gender\_id, user\_age\_level, user\_star\_level}\} for public dataset, and \{\texttt{scenario, ad\_position, user\_profile}\} for production dataset.} We split each into training/dev/test sets by timestamp with 4:1:1 proportion.

\subsubsection{\textbf{Competitors}} We use \textsf{DNN} ~\cite{covington2016deep} and \textsf{DCNv2}~\cite{wang2021dcn} as two backbones of prediction models. On this basis, we compare \textsf{AdaSparse} with the previous state-of-the-art multi-domain CTR prediction approaches: 
(1) \textsf{MAML}~\cite{yu2020personalized} treats each domain as a meta-task and  quickly learns domain-specific parameters from a small amount of domain data. 
(2) \textsf{STAR}~\cite{sheng2021one} learns predefined domain-specific parameters using the data from this domain only. 
(3) \textsf{APG}~\cite{yan2022apg} generates domain-specific parameters using additional networks, and uses low-rank decomposition to reduce computation cost.\footnote{Note that for the approaches based on individual parameters, they cannot be used for thousands of domains due to huge memory cost. Therefore, we retain a most representative feature in the domain-aware feature set.} 

The metrics of LogLoss, AUC and Group AUC (GAUC~\cite{he2016ups} for short) of domains are used for evaluation. For fair comparison, all approaches have the same feature embedding size, and the main network contains three layers. The output sizes of hidden layers are set to \{512,256,128\} (Production) and \{128,64,32\} (Public). We use Adam optimizer with 1024 (Production) and 256 (Public) batch size and 0.001 learning rate. For the pruner, a fully-connected layer is used to produce weight factors of each layer $l$. For each comparative model we carefully tune their hyperparameters. The hyper-parameters in \textsf{AdaSparse} are set as the following: $\beta=2$, $\epsilon=0.25$, $\alpha$ is initialized to 0.1 and its upper limit is set to 5, $r_{min}=0.15, r_{max}=0.25$.

\subsection{Results and Discussion}
\subsubsection{\textbf{Main Results}} Table~\ref{results:main} shows the results of multi-domain CTR predictive approaches, with \textsf{DNN} and \textsf{DCNv2} as backbone respectively. On both datasets, all multi-domain methods achieve better performance compared to \textsf{DNN} or \textsf{DCNv2}, demonstrating that capturing the differentiated user preferences for multiple domains is the key for CTR prediction. 
Compared to \textsf{DNN}, \textsf{DCNv2} performs better on Production dataset but worse on Public dataset. We empirically explain that because the size of Public dataset is too small, it is hard to train \textsf{DCNv2} (with more parameters) adequately. 

\textsf{APG} and \textsf{AdaSparse} outperform \textsf{MAML} and \textsf{STAR}, verifying that approaches based on generated parameters have advantages over those based on domain individual parameters. Benefiting from parameter generating and pruning redundant neurons, \textsf{AdaSparse} learns domain-adaptive sparse structures, achieving a significant improvement compared to all competitors.

\begin{table}[t]
\footnotesize
\caption{Results on multi-domain CTR prediction. ``*'' denotes that \textsf{AdaSparse} significantly outperforms the second-best approach at the level of $p<0.05$. }
\centering
\setlength{\tabcolsep}{0.8mm}{
\begin{tabular}{lcccccc}
\toprule
\multirow{2}*{\textbf{Approach}} & \multicolumn{3}{c}{\textbf{Production Dataset}}&
\multicolumn{3}{c}{\textbf{Public Dataset}} \\
\cmidrule(lr){2-4}\cmidrule(lr){5-7}
 & LogLoss & AUC &  GAUC & LogLoss & AUC &  GAUC  \\   
\midrule
\textsf{DNN} & 0.063630 & 0.7266 & 0.6669 & 0.094886& 0.6506  & 0.6442 \\
\textsf{\qquad+MAML}& 0.063474 &0.7308 & 0.6715 & 0.093912 & 0.6531 & 0.6463 \\
\textsf{\qquad+STAR} & 0.063445 & 0.7313 & 0.6762 & 0.093598 & 0.6541 & 0.6505\\
\textsf{\qquad+APG} & 0.063232 &0.7332 & 0.6826 & 0.093264 & 0.6585 & 0.6553 \\
\cmidrule(lr){1-1}
\textsf{\qquad+AdaSparse}& \textbf{0.063215} & \textbf{0.7359}$^*$ & \textbf{0.6848}$^*$  & \textbf{0.093139} &  \textbf{0.6607}$^*$   & \textbf{0.6572}$^*$ \\
\cmidrule(lr){1-7}
\textsf{DCNv2} & 0.063602 & 0.7282 & 0.6684 & 0.095097 &0.6497  & 0.6443 \\
\textsf{\qquad+MAML}& 0.063619 & 0.7326 & 0.6730 & 0.094024 & 0.6536 & 0.6474 \\
\textsf{\qquad+STAR}& 0.063424 & 0.7325 & 0.6794 & 0.093754 & 0.6538 & 0.6501 \\
\textsf{\qquad+APG} & 0.063192 & 0.7356 & 0.6840 & 0.093375 & 0.6572 & 0.6536 \\
\cmidrule(lr){1-1}
\textsf{\qquad+AdaSparse} & \textbf{0.063187} & \textbf{0.7378}$^*$ & \textbf{0.6871}$^*$ & \textbf{0.093292} & \textbf{0.6594}$^*$ & \textbf{0.6565}$^*$  \\
\bottomrule
\end{tabular}}
\label{results:main}
\end{table}

\begin{table}[t]
\footnotesize
\centering
\caption{Comparison of different weighting factors.}
\begin{tabular}{lcc}
\toprule
\textbf{Methods} & AUC($\uparrow$) & GAUC($\uparrow$) \\ 
\midrule
{\textsf{DNN}} & {0.7266} & {0.6669} \\
{\textsf{\qquad+ AdaSparse (Binarization)}} & \underline{0.7355} & \underline{0.6843} \\
{\textsf{\qquad+ AdaSparse (Scaling)}} & {0.7334} & {0.6817}  \\
{\textsf{\qquad+ AdaSparse (Fusion)}} & \textbf{{0.7359}} & \textbf{{0.6848}}  \\
\bottomrule
\end{tabular}
\label{table:discussion_factors}
\end{table}

\subsubsection{\textbf{Comparison among Three Formulations of Factors}} Table~\ref{table:discussion_factors} shows the performance of three weighting factors based on \textsf{DNN} backbone. 
The effectiveness of {Binarization} method verifies that pruning the redundant information specific to each domain can improve the overall generalization. 
{Scaling} method considers the importance of neurons and also boosts the performance. 
{Fusion} method that combines {Binarization} and {Scaling} achieves the best performance. This shows that taking both redundance and importance of neurons into account is indeed an effective strategy for improving multi-domain CTR prediction.

\subsection{Further Analysis}

\subsubsection{\textbf{Generalization across Domains}} To explicitly understand \textsf{AdaSparse}'s ability of capturing the characteristics of domains, we select three domains' subset: denoted as $Da$, $\widehat{Da}$ and $Db$, where the data distributions of $Da$ and $\widehat{Da}$ are similar, while $Db$ is quite different from them. Fig.~\ref{fig:futher_study} (a) visualizes the learned binary weighting factors for the last 128-dimension layer. The gray areas represent 0 value and others mean 1 value. As we can see, $Da$ and $\widehat{Da}$ have a great proportion of overlaps, while $Db$ only has a small proportion. This shows that \textsf{AdaSparse} can capture both the commonalities and characteristics among domains.

To verify the performance boosts from top to long-tail domains, we rank all 5,000+ domains by their sample sizes, and then merge them into 5 bins based on equal-frequency segmentation. 
Let \textit{BIN1 / BIN5} denote the top/long-tailed domains respectively, which contain \textit{18 / 3000+ domains} with both $20\%$ of samples. 
As shown in Fig.~\ref{fig:futher_study} (b), all approaches improve significant performance on top domains (BIN1, 2 and 3). This verifies that rich training data is helpful for learning domain-specific parameters. 
However if we consider long-tail domains (BIN4 and 5), only \textsf{AdaSparse} and \textsf{MAML} can still achieve improvement by a large margin. 
Benefiting from pruning domain-aware redundancy, \textsf{AdaSparse} achieves better performance long-tail domains having limited training data, and has the best generalization across all domains.

\subsubsection{\textbf{Effect of Different Domain Partitions}}
To analyze the impact of domain partitions, we modify the domain-aware feature set. We first partition the dataset into 10 domains using the feature \texttt{scenario}. Then, we gradually add new features~\footnote{We add \{ \texttt{scenario}, \texttt{ad\_position}, \texttt{user\_profile}, \texttt{user\_behaviors}\} sequentially.} to divide the dataset into more domains. 
As shown in Fig.~\ref{fig:futher_study} (c), \textsf{STAR} has a slight advantage over \textsf{AdaSparse} under a small number of domains. \textsf{AdaSparse}, however, achieves significant AUC boosts as the number of domains increases. This suggests that the effectiveness may be enhanced by using fine-grained domain partitioning. Additionally, the results also demonstrate that having too many partitions will reduce the effectiveness.

\subsubsection{\textbf{Sensitiveness of Sparsity Controlling for Pruning}} 
\begin{figure}[t]
\centering
\centerline{\includegraphics[width=\columnwidth]{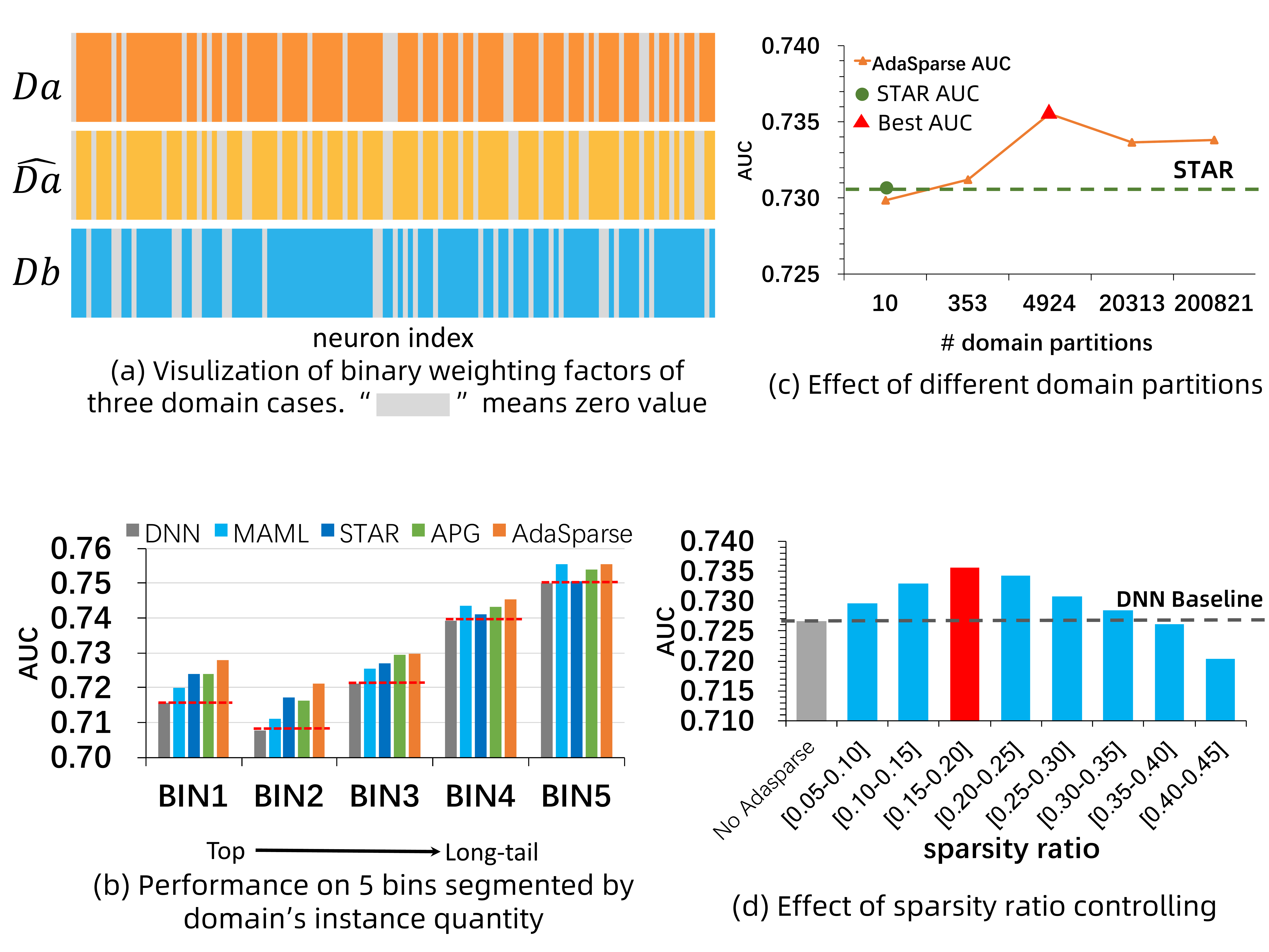}}
\caption{(a) visualizes the binary weighting factors from  two similar domains and one discrepant domain. (b) shows the different performance boosts from top to long-tail domains. (c) verifies that fine-grained domain division improves generalization. (d) shows the effect of sparsity ratio controlling.}
\label{fig:futher_study}
\end{figure}

We change the hyperparameters \textit{sparsity ratio boundary} $[r_{min},r_{max}]$ for evaluation. 
Here, we empirically consider that a model does not require a sparsity ratio higher than $0.5$. Therefore, we test several groups of sparsity ratios that are all less than $0.5$. As seen in Fig.~\ref{fig:futher_study} (d), the performance grows with increasing sparsity ratio and ultimately declines. This supports the idea that pruning redundant information can improve effectiveness. We also observe that even if we use $[0.35,0.40]$ as the sparsity ratio boundary, the resulting model performs nearly as well as the base \textsf{DNN} model. This indicates that the model structure is usually over-parameterized and contains a high proportion of redundant parameters.

\subsection{Online A/B Test}
We conduct online experiments in our advertising system's CTR module for 10 days. 
We use metrics including $CTR=\frac{\text{\#click}}{\text{\#impression}}$, $CPC= \frac{\text{cost of advertisers}}{\text{\#click}}$, and define a new metric named uplifted domains ratio ($UPR = \frac{\text{ \# Uplifted domains }}{\text{\#Total domains}}$) to show the proportion of domains having performance boost.  \textsf{AdaSparse} brings \textbf{4.63}\% gain on CTR and \textbf{3.82}\% reduction on CPC, and improves the CTR on $90\%+$ of domains, verifying its domain-effectiveness in industry.

\section{Conclusion}
We propose \textsf{AdaSparse} to learn adaptively sparse structures for multi-domain  CTR prediction. It prunes redundant neurons w.r.t different domains via learned neuron-level weighting factors to improve generalization. Both offline and online experiments verify that \textsf{AdaSparse} outperforms previous models~\cite{yan2022apg,sheng2021one,yu2020personalized}. 
In future work we will combine pruning and neural architecture search techniques~\cite{wei2021autoheri} to further improve generalization across domains.

\bibliographystyle{ACM-Reference-Format}
\bibliography{research_src-adasparse.bib}

\end{document}